\documentclass[conference]{IEEEtran}
\IEEEoverridecommandlockouts

\usepackage{cite}
\usepackage{amsmath,amssymb,amsfonts}
\usepackage{algorithmic}
\usepackage{graphicx}
\usepackage{textcomp}
\usepackage{multirow}
\usepackage{verbatim}
\usepackage{color}
\usepackage{xcolor}
\definecolor{hiddendraw}{rgb}{0.7,0.7,0.7}
\usepackage{tikz}
\usepackage[edges]{forest}
\def\BibTeX{{\rm B\kern-.05em{\sc i\kern-.025em b}\kern-.08em
    T\kern-.1667em\lower.7ex\hbox{E}\kern-.125emX}}
\begin{document}

\title{A Survey of Medical Point Cloud Shape Learning: Registration, Reconstruction and Variation\\
}

\author{\IEEEauthorblockN{Tongxu Zhang\textsuperscript{1, 2}, Zhiming Liang\textsuperscript{1}, Bei Wang\textsuperscript{1}}
\IEEEauthorblockA{\textit{\textsuperscript{1} East China University of Science and Technology, Xuhui, Shanghai, China} \\
\textit{\textsuperscript{2} The Hong Kong Polytechnic University, Kowloon, Hong Kong SAR, China} \\
$\{$juk, y80230193$\}$@mail.ecust.edu.cn, beiwang@ecust.edu.cn}
}

\maketitle

\begin{abstract}
Point clouds have become an increasingly important representation for 3D medical imaging, offering a compact, surface-preserving alternative to traditional voxel or mesh-based approaches. Recent advances in deep learning have enabled rapid progress in extracting, modeling, and analyzing anatomical shapes directly from point cloud data. This paper provides a comprehensive and systematic survey of learning-based shape analysis for medical point clouds, focusing on three fundamental tasks: registration, reconstruction, and variation modeling. We review recent literature from 2021 to 2025, summarize representative methods, datasets, and evaluation metrics, and highlight clinical applications and unique challenges in the medical domain. Key trends include the integration of hybrid representations, large-scale self-supervised models, and generative techniques. We also discuss current limitations, such as data scarcity, inter-patient variability, and the need for interpretable and robust solutions for clinical deployment. Finally, future directions are outlined for advancing point cloud-based shape learning in medical imaging.
\end{abstract}


\section{Introduction}
3D data can usually be represented in various formats, including depth images, point clouds, meshes, and volumetric grids. As a commonly used format, point clouds preserve the original geometric information in 3D space without any discretization. With the advancement of 3D scanning technologies such as Computed Tomography (CT) and Magnetic Resonance Imaging (MRI) \cite{doi2007computer}, 3D point clouds are increasingly adopted as a way to represent complex anatomical structures. The development of post-processing techniques for volumetric CT and MRI has made converting volumetric images into point clouds as a lightweight representation a common practice—this provides a lightweight, topology-independent spatial sampling set that retains millimeter-level detail while avoiding the memory overhead of meshes or dense grids \cite{zhang2024deepsurvey}.

\begin{figure*}[htbp]
\centering
\includegraphics[width=\textwidth]{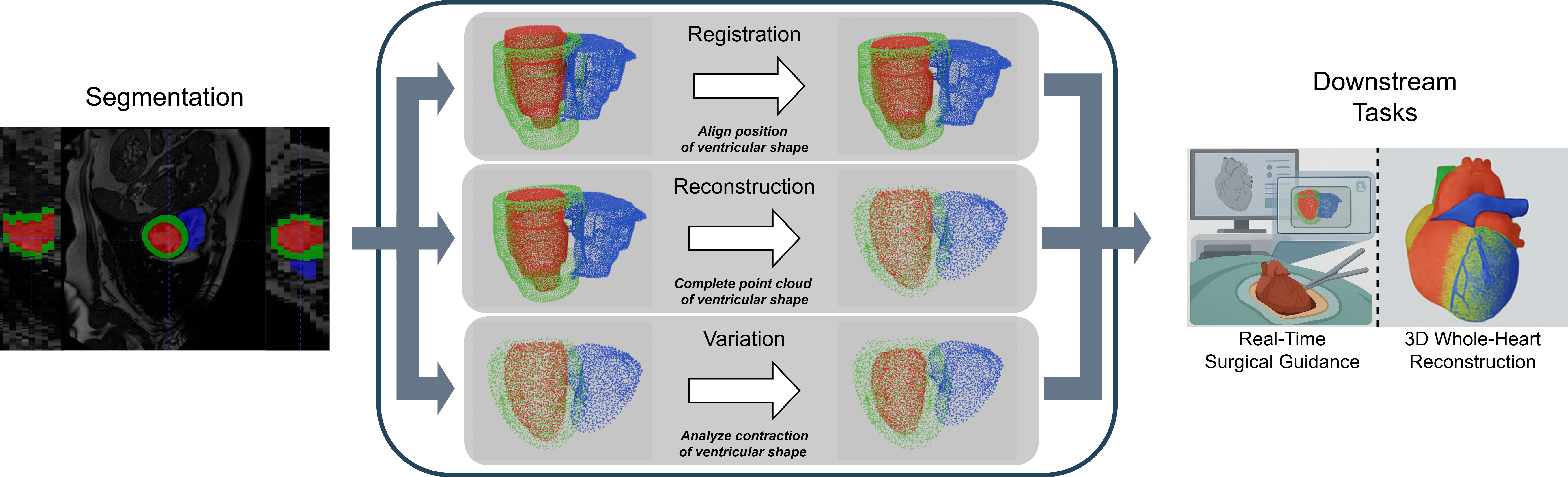}
\caption{Overview of the shape learning pipeline for medical point clouds. The process begins with segmentation, which extracts anatomical structures from imaging data. Three core learning-based tasks—registration, reconstruction, and variation—are then performed after segmentation: registration aligns shapes, reconstruction completes partial point clouds, and variation analyzes morphological differences. The outputs from these tasks support a range of downstream clinical and analytical applications.}
\label{fig:flow}
\end{figure*}

Recent advances in deep learning have accelerated 3D medical image analysis, especially for shape learning with point clouds~\cite{singh20203d,chen2019med3d}. While voxel-based representations remain prevalent~\cite{cciccek20163dunet,wang2025sammed3d}, point clouds offer a compact, surface-preserving, and anatomically aligned alternative that enables efficient shape understanding~\cite{yu20213d,bruse2016statistical}.


In this context, Shape Learning refers to extracting, encoding, and manipulating anatomical geometry from point cloud data through learning-based methods~\cite{gutierrez2019learning,adams2023can,adams2024point2ssm++}. Accordingly, there are three fundamental downstream tasks: registration, reconstruction, and shape variation. Segmentation serves as the prerequisite for these tasks, providing essential anatomical priors that enable accurate registration, robust reconstruction, and meaningful variation modeling. Thus, segmentation and the three shape learning tasks form a progressive pipeline, rather than parallel processes. Meanwhile, classification, prediction~\cite{choi2024deep}, and morphometrics~\cite{zhu2025point} often rely on high-quality registration, reconstruction, and variation outcomes. Specifically, registration aligns anatomical structures across modalities, time points, or subjects, enabling change detection, atlas construction, and longitudinal analysis. Conversely, reconstruction aims to recover complete anatomical geometry from sparse or partial observations, including tasks such as completion, upsampling, and implicit surface modeling (here, this also includes segmentation for morphology understanding). Shape modeling focuses on learning shape distributions at both individual and population levels, supporting multiple tasks such as reconstruction, identification, and morphological analysis. The overall pipeline and relationships between segmentation, the three shape learning tasks, and their downstream applications are illustrated in Fig.~\ref{fig:flowchat}.

Classical methods such as Iterative Closest Point (ICP) for rigid alignment and B-spline-based Demons algorithms for non-rigid registration have been widely applied in the medical field \cite{thirion1998image,wang2023bone}. However, these methods often prove fragile when faced with sparse sampling, missing data, or multimodal discrepancies. Recent advancements in deep learning for point clouds, including PointNet \cite{qi2017pointnet}, DGCNN \cite{wang2019dynamic}, and Transformer-based architectures \cite{zhao2021point}, have introduced new paradigms for Shape Learning. These methods can learn hierarchical shape descriptors, infer deformation fields, or reconstruct organ surfaces from scattered and noisy inputs. Nevertheless, adapting these methods to the medical domain presents unique challenges, including limited annotated data, anatomical variability, cross-modal inconsistencies, and the demand for interpretability and reliability in clinical workflows.

Several recent studies have started addressing these challenges. For instance, Hansen and Heinrich \cite{hansen2021deep} significantly outperformed dense CNN-based registration in lung alignment for Chronic Obstructive Pulmonary Disease (COPD) patients by employing graph-based geometric reasoning and Loopy BP. PCD-Net \cite{beetz2024modeling} achieved sub-voxel deformation learning for cardiac MRI point clouds. In the field of reconstruction, methods like MedShapeNet \cite{li2025medshapenet,yassin2024medshapenet} and MedPointS \cite{zhang2025hierarchical} have demonstrated the potential of large-scale datasets for medical and anatomical classification, completion, and segmentation in point cloud completion tasks.

Despite these advancements, there remains a lack of a unified comparative overview of Shape Learning methods, particularly those focusing on Registration, Reconstruction, and Variation from medical point clouds. Existing surveys either concentrate on general point cloud processing \cite{guo2020deep} or emphasize image-based learning pipelines, with limited attention to anatomical priors and clinical constraints. This paper is the first comprehensive review covering the application of Shape Learning in medical imaging. This survey aims to fill this gap by providing a comprehensive and comparative overview of the latest developments in medical point cloud Registration, Reconstruction, and Variation, with a focus on learning-based approaches. Specifically:

1. We conduct a systematic and comprehensive review of the work on medical point cloud Registration, Reconstruction, and Variation over the past five years, analyzing and comparing state-of-the-art methods for each specific task.

2. We summarize the most advanced methods for Registration, Reconstruction, and Variation, highlighting their models, datasets, and loss functions.

3. We discuss the challenges and unresolved issues related to point cloud-based Shape Learning in medical imaging. We identify new trends, propose key questions, and suggest future directions for further exploration.

\section{Literature Collection}

\begin{figure}[htbp]
\centerline{\includegraphics[scale=0.13]{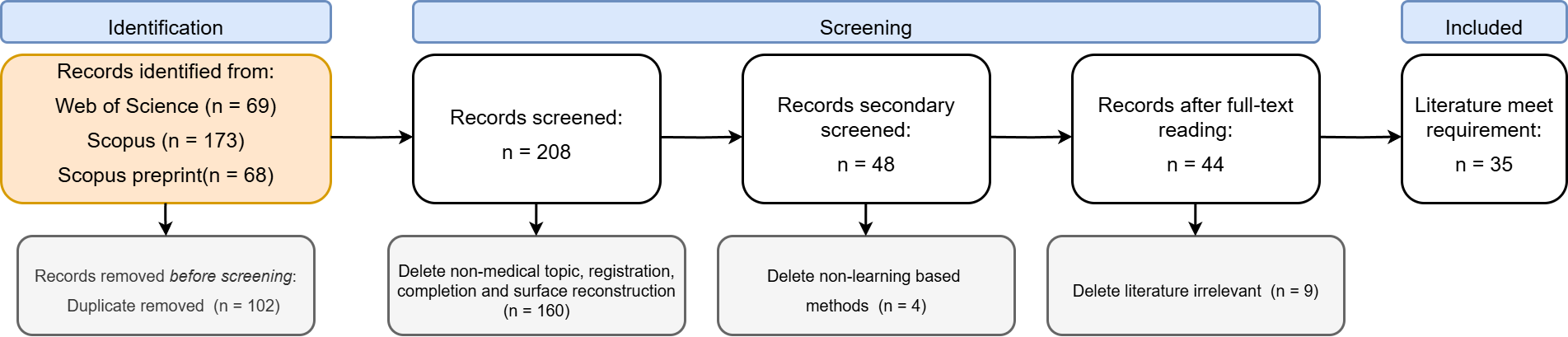}}
\caption{Literature selection based on the PRISMA guideline. Records were identified from WOS, Scopus, and preprints, then screened by title, abstract, and full text. After removing duplicates and irrelevant works, 35 publications were finally included for comparative review.}
\label{fig:prisma}
\end{figure}

Following the PRISMA guideline~\cite{Pagen71}, we systematically searched Web of Science (WOS) and Scopus, including preprints, for publications from 2021 to 2025, using keywords such as ("point cloud" AND "medical" AND "tasks") and ("point cloud" AND "medical image"). After screening and deduplication, 35 relevant studies on Registration, Reconstruction, and Variation, with particular attention to radiological images such as CT, MRI, PET, and SPECT, focusing on medical point clouds were included in this survey, as shown in Fig.~\ref{fig:prisma}. 

\tikzstyle{leaf}=[draw=hiddendraw,
    rounded corners,minimum height=1em,
    fill=orange!40,text opacity=1, align=center,
    fill opacity=.5,  text=black,align=left,font=\scriptsize,
    inner xsep=3pt,
    inner ysep=2.5pt,
    ]
\begin{figure*}[ht]
\centering
\begin{forest}
  for tree={
    forked edges,
    grow=east,
    reversed=true,
    anchor=base west,
    parent anchor=east,
    child anchor=west,
    base=middle,
    font=\scriptsize,
    rectangle,
    draw=hiddendraw,
    rounded corners,align=left,
    minimum width=2em,
    s sep=5pt,
    inner xsep=3pt,
    inner ysep=2.5pt,
  },
  where level=1{text width=5em,font=\scriptsize}{},
  where level=2{text width=7em,font=\scriptsize}{},
  where level=3{text width=9em,font=\scriptsize}{},
  where level=4{font=\scriptsize}{},
  where level=5{font=\scriptsize}{},
  [Shape Learning Methods for Medical Point Clouds,rotate=90,anchor=north,edge=hiddendraw
    [Registration, text width=4.5em, align=center
      [Rigid, text width=8em
        [Adaptive Super4PCS~\cite{su20213d}{,} Bastico et al.~\cite{bastico2024coupled}{,} Heiselman et al.~\cite{heiselman2024image}, leaf, text width=31.5em]
      ]
      [Deformable, text width=8em
        [Mekhzoum et al.~\cite{mekhzoum2024towards}{,} Joutard et al.~\cite{joutard2022multi}{,} Hansen \& Heinrich~\cite{hansen2021deep}{,} LCNet~\cite{liu2025lcnet}, leaf, text width=31.5em]
      ]
      [Cross-modal, text width=8em
        [Wang et al.~\cite{wang2023bone}{,} Falta et al.~\cite{falta2024unleashing}{,} Zhang et al.~\cite{zhang2023volumetric}{,} PointVoxelFormer~\cite{heinrich2024pointvoxelformer}, leaf, text width=31.5em]
      ]
    ]
    [Reconstruction, text width=4.5em, align=center
      [Point Cloud Completion, text width=8em
        [SA-PoinTr~\cite{chen2024cartilage}{,} MSN-FM~\cite{yassin2024medshapenet}{,} Flemme~\cite{zhang2025hierarchical}{,} Hu et al.~\cite{hu2021point,hu2022srt}{,} Beetz et al.~\cite{beetz2023multi}{,} Mazzocchetti et al.~\cite{mazzocchetti2024neural},leaf, text width=31.5em]
      ]
      [Surface Reconstruction, text width=8em
        [Xiong et al.~\cite{xiong2022automatic}{,} Point2Mesh-Net~\cite{beetz2022point2mesh}{,} PointNeuron~\cite{zhao2023pointneuron}{,} CPAConv-POCO~\cite{jiang2023cpaconv}, leaf, text width=31.5em]
      ]
      [Completion \& Surface, text width=8em
        [MR-Net~\cite{chen2021shape}{,} Zhang et al.~\cite{xzhang2023anatomical,xzhang2024robust}{,} Gutiérrez-Becker et al.~\cite{gutierrez2021discriminative}, leaf, text width=31.5em]
      ]
    ]
    [Variation, text width=4.5em, align=center
      [Statistical Shape Model, text width=8em
        [PUSSM~\cite{zhang2025pussmpointcloudupsampling}{,} DeepSSM~\cite{bhalodia2024deepssm}{,} BVIB-DeepSSM~\cite{adams2024weakly}{,} Point2SSM++~\cite{adams2024point2ssm++}{,} Mesh2SSM~\cite{iyer2023mesh2ssm} \\ Adams \& Elhabian~\cite{adams2023can},leaf, text width=31.5em]
        ]
      [Signed Distance Function, text width=8em
        [Cardiac DeepSDF~\cite{verhulsdonk2023shape}, leaf, text width=31.5em]
      ]
      [Application, text width=8em
        [PCD-Net~\cite{beetz2024modeling}{,} Zhang Q. et al.~\cite{zhang2024deformation}, leaf, text width=31.5em]
      ]
    ]
  ]
\end{forest}
\caption{A taxonomy of shape learning-based methods for medical point clouds with representative works.}
\label{fig:taxonomy_shapelearning}
\end{figure*}
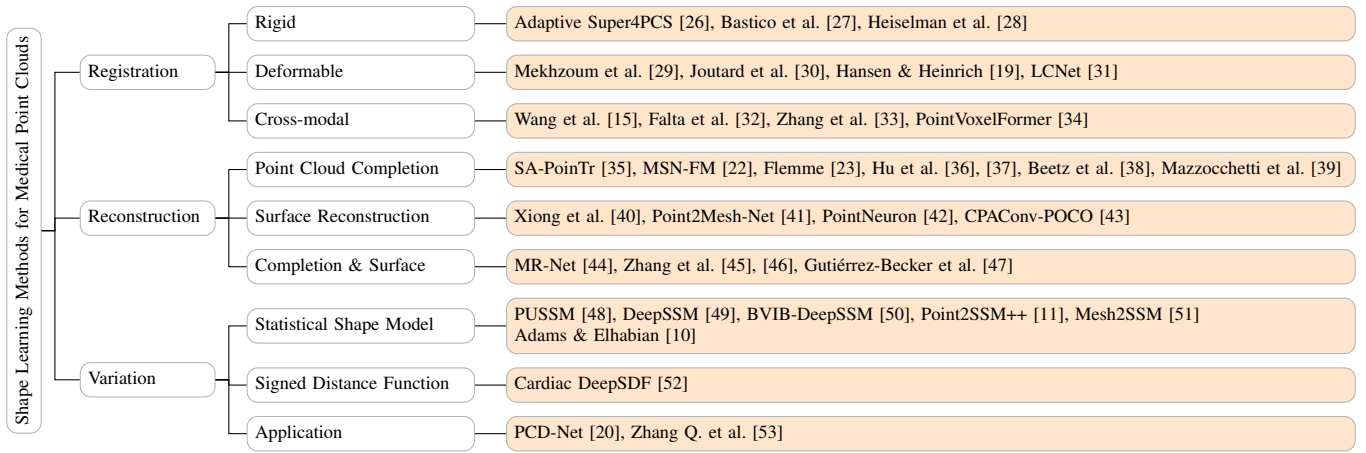

\section{Clinical Importance}
Modern medical decision-making increasingly relies on accurate and efficient 3D anatomical structure representation. This paper focuses on two major tasks—Registration and Reconstruction—using point clouds as the basic data form, with their clinical importance reflected in the following four aspects:

\subsubsection{Improving the Accuracy of Surgical Planning and Navigation}
Effectively learning and processing shapes within medical point clouds has diverse and critical applications in clinical practice. These applications include improving diagnosis through precise anatomical quantification and providing accurate 3D models for simulation and navigation-guided surgical operations~\cite{henrich2024ludo,acar2025monocular,lin2023surface}. Registration ensures spatial consistency between preoperative analysis and intraoperative navigation.

\subsubsection{Supporting Visualization and Patient Communication}
The 3D shapes recovered by reconstruction from sparse or occluded data can provide intuitive visual support for both patients and doctors. Consistent comparison and analysis of anatomical changes over time are enabled, facilitating multimodal image fusion and motion correction studies~\cite{yu2023motility,zhao2023novel}. This is especially useful for orthopedic reconstruction (such as craniofacial implants), where automated stitching of craniofacial defects~\cite{mazzocchetti2024neural} assists in customizing implants. However, it should be noted that point clouds, while effective at encoding surface geometry, may have limited ability to capture fine-grained internal tissue structures.

\subsubsection{Reducing Radiation Dose and Acquisition Costs}
Reconstruction using point clouds can restore shapes from a small number of image slices or surface scans. In interventional radiotherapy and cardiac catheterization, accurate registration significantly reduces errors, thereby improving the identification of treatment target areas~\cite{hansen2022phd}, markedly lowering radiation risks. It can also be used for rapid 3D structure modeling based solely on intraoperative point clouds, such as atrial mapping in atrial fibrillation ablation procedures~\cite{chen2021shape}.

\subsubsection{Data Privacy Friendly and Highly Generalizable Across Centers}
Although point clouds still preserve morphological features, training shape data encourages deep networks to focus on learning discriminative geometric characteristics rather than patient identity information irrelevant to the task~\cite{li2025medshapenet}. This helps improve the robustness and reliability of learning systems and prevents identity-driven biases. Nevertheless, if point cloud density is sufficiently high, there remains a theoretical risk of re-identifying individual patients; thus, privacy-preserving strategies should still be considered in practical applications.

\section{Taxonomy}
To provide a systematic overview, shape learning-based methods for medical point clouds can be organized into three primary categories: registration, reconstruction, and variation. Each category includes several representative sub-tasks and methods, as illustrated in Fig.~\ref{fig:taxonomy_shapelearning}.
\subsection{Registration}
Registration of medical point clouds aims to align 3D shapes across time points or imaging modalities, supporting tasks such as shape comparison, surgical planning, and reconstruction. Methods have evolved from classical iterative closest point (ICP) variants to advanced deep learning and generative models, typically categorized as rigid, deformable, and cross-modal registration.

For rigid registration, recent work has improved robustness and automation. Adaptive Super4PCS~\cite{su20213d} removes manual threshold tuning for stable multi-device cranial alignment. The coupled Laplacian approach~\cite{bastico2024coupled} enhances matching accuracy under anatomical variation, while Heiselman et al.~\cite{heiselman2024image} provide a comprehensive benchmark of classic, deep learning, and FEM-based methods on sparse liver data.

Deformable registration has shifted toward geometric deep learning frameworks for complex anatomical variation. Graph-ICP~\cite{hansen2021deep} leverages geometric features and loopy belief propagation, reducing TRE and improving resilience to artifacts. Mekhzoum et al.~\cite{mekhzoum2024towards} apply deep point descriptors to dynamic 4D-CT, and Joutard et al.~\cite{joutard2022multi} extend Bayesian Coherent Point Drift to model organ elasticity and inter-organ coherence. LCNet~\cite{liu2025lcnet} further enables robust intraoperative liver deformation correction.

Cross-modal registration tackles alignment across modalities such as CT, MRI, and CBCT, facing challenges from appearance and resolution differences. CSN-ICP~\cite{wang2023bone} couples nearest neighbor search in Cartesian and spherical spaces to reduce mismatches in sparse bone clouds. Diffusion-based synthetic data~\cite{falta2024unleashing} mitigates paired data scarcity, while keypoint-guided methods~\cite{zhang2023volumetric} and hybrid frameworks like PointVoxelFormer~\cite{heinrich2024pointvoxelformer} enable efficient deformation learning in large-scale and multi-modal scenarios.

Despite progress, deformable and cross-modal registration remain more challenging than rigid cases, due to non-rigid object complexity and reliance on annotated or paired data. Promising directions include weakly-supervised/unsupervised deformation learning, synthetic data for bridging modality gaps, and end-to-end pipelines combining registration and reconstruction for improved consistency and completeness.

\subsection{Reconstruction}

Medical point cloud reconstruction has rapidly evolved in recent years, spanning key tasks such as point cloud completion, surface (mesh) reconstruction, anatomical partitioning, and topological recovery. Rather than simple geometric filling, modern approaches increasingly focus on capturing anatomical fidelity and clinical applicability, leveraging the progress of deep learning, hybrid representations, and large-scale data resources.

A prominent trend is the adoption of transformer-based and self-supervised models, which have greatly advanced generalization and adaptability for complex and sparse medical data. For example, SA-PoinTr~\cite{chen2024cartilage} introduced vision transformer concepts into point cloud completion, using multi-branch self-attention to effectively capture local anatomical variation. This method achieved over 6\% F-score improvement on challenging femoral cartilage defect datasets, and avoided mesh holes or redundancies commonly observed in previous approaches. Similarly, MSN-FM~\cite{yassin2024medshapenet} applied GPT-like priors for medical shapes, pretraining on 200,000 multi-organ shapes and incorporating textual labels for zero- and few-shot completion, which proved robust to non-uniform point sampling. Building on the direction of large-scale benchmarks, Flemme~\cite{zhang2025hierarchical} introduced a state space model for efficient completion of complex nested structures such as the colon, outperforming previous methods and establishing the MedPointS dataset with nearly 29,000 point clouds.

Innovations are not limited to network architecture: the integration of domain priors and multi-scale representations is equally crucial. Point-voxel fusion frameworks, such as those by Zhang et al.~\cite{xzhang2023anatomical,xzhang2024robust}, were among the first to jointly model vessel structure priors and continuous spatial sampling. This enabled multi-scale feature fusion, resolving Couinaud segment boundary ambiguities and generating smooth anatomical planes, with average surface distance reduced to 3.19mm. Meanwhile, Beetz et al.~\cite{beetz2023multi} combined cardiac point cloud reconstruction with myocardial infarction prediction, constraining shape features via KL-divergence and improving both reconstruction precision (CD$<$1.71mm) and clinical prediction (AUC increased by 19\%).

Surface reconstruction and topology learning have also seen notable progress. Point2Mesh-Net~\cite{beetz2022point2mesh} innovatively combined PointNet and graph convolutional upsamplers to produce topologically consistent cardiac meshes in a single pass, demonstrating robustness to slice misalignment and sparsity—features that support integration into large clinical biobanks. CPAConv-POCO~\cite{jiang2023cpaconv} was the first to introduce irregular point cloud convolution into occupancy networks, substantially improving lung nodule reconstruction accuracy and showing organ-agnostic generalization even for topologically complex lesions. PointNeuron~\cite{zhao2023pointneuron} pioneered the reformulation of neuron structure reconstruction as a joint geometry-topology learning problem, employing GCN for skeleton proposals and graph auto-encoders for connectivity and explicit radius prediction, enhancing the representation of skeletal structures. For rapid intraoperative integration, Xiong et al.~\cite{xiong2022automatic} employed large-kernel 3D CNNs to learn sparse point cloud shells in voxel space, though their approach remains sensitive to noise and requires manual registration.

Some methods, like MR-Net~\cite{chen2021shape}, blend point cloud completion with mesh surface deformation: using 3D CNNs to extract contour voxel features and GCNs to deform template meshes, achieving robust performance even with several missing imaging slices. Graph-based and generative approaches, such as Hu et al.~\cite{hu2021point,hu2022srt}, advanced real-time intraoperative brain reconstruction through GAN and Transformer architectures, reducing reconstruction errors by over 25\% compared to previous methods. Gutiérrez-Becker et al.~\cite{gutierrez2021discriminative} proposed a conditional variational autoencoder for unordered point clouds, enabling robust shape generation and analysis with disease label conditioning—crucially, without the need for point correspondence or pre-aligned topology.  Notably, Mazzocchetti et al.~\cite{mazzocchetti2024neural} introduced a new dataset and comparative study for maxillofacial shape completion, identifying state-of-the-art methods that best meet clinical demands.

Overall, the field is witnessing several converging research directions: 1. Hybrid point-voxel structures and multimodal fusion are becoming mainstream, effectively balancing computational efficiency and anatomical precision, and improving model robustness under sparse or defective conditions; 2. Large model self-supervised learning and knowledge-driven pretraining, such as those built upon MedShapeNet~\cite{li2025medshapenet}, are enabling small-sample and cross-task transfer, lowering the barrier for customized shape models; 3. Application scenarios are rapidly expanding from single-organ and simple surface tasks to complex, multi-organ, fine-partition reconstruction, as well as unified geometric-topological-semantic learning—together, these trends are paving the way for advanced applications in AR/VR surgical simulation and medical education.

\subsection{Variation}

In medical point cloud learning, the study of Variation focuses on capturing, modeling, and analyzing the morphological diversity of organs and anatomical structures—both across populations and at the individual level. This line of research is crucial for understanding the latent shape space that encodes anatomical variability, which, in turn, supports downstream applications such as disease assessment, surgical planning, and morphometric analysis. Unlike Registration and Reconstruction, which emphasize alignment and geometric recovery, Variation is concerned with learning distributions, measuring deviations, and embedding anatomical priors into robust shape models.

A central paradigm in this field is Statistical Shape Modeling (SSM). Traditional SSMs, such as Point Distribution Models (PDM), leverage principal component analysis (PCA) to capture the principal axes of shape variation from collections of 3D objects. Early work by Adams \& Elhabian~\cite{adams2023can} demonstrated that point clouds can be directly employed as input for PDMs, making use of the accessibility and flexibility of point-based shape representations in clinical imaging pipelines. Building on this, more recent frameworks, including DeepSSM~\cite{bhalodia2024deepssm}, have proposed end-to-end learning pipelines that map medical images directly to statistical shape descriptors and correspondence fields, automating the analysis of population-level anatomical variability. Extensions such as Point2SSM++~\cite{adams2024point2ssm++} and Mesh2SSM~\cite{iyer2023mesh2ssm} explicitly construct SSMs by learning correspondences, supporting multi-organ and even four-dimensional (dynamic) morphometric analysis while minimizing template bias. The PUSSM model~\cite{zhang2025pussmpointcloudupsampling} adds another dimension by introducing implicit statistical priors via upsampling networks, densifying segmentation-derived point clouds and thus improving downstream reconstruction fidelity.

Beyond explicit statistical modeling, several studies have embraced implicit and generative approaches. For example, Cardiac DeepSDF~\cite{verhulsdonk2023shape} employs deep signed distance functions (DeepSDF) to efficiently parameterize the underlying shape space of the heart, enabling robust completion and modality transfer even from highly sparse input points. In parallel, methods like BVIB-DeepSSM~\cite{adams2024weakly} have explored weakly supervised probabilistic shape modeling, demonstrating that end-to-end latent space learning can scale to large, heterogeneous datasets with limited annotations, thus facilitating clinical translation.

Deformation field modeling is another critical facet of the Variation landscape, especially in settings requiring biomechanical interpretability or compensation for intraoperative anatomical changes. PCD-Net~\cite{beetz2024modeling} is notable for capturing dynamic, disease-specific cardiac deformation fields, supporting myocardial infarction (MI) detection and prognosis with strong interpretability. Similarly, Zhang Q. et al.~\cite{zhang2024deformation} successfully predicted liver deformation fields from ultra-sparse intraoperative point clouds, offering promising avenues for surgical navigation and compensation. The importance of template bias and learning efficiency is further highlighted by works like Mesh2SSM~\cite{iyer2023mesh2ssm}, which uses self-learning templates for rapid, low-bias inference.

Together, these advances highlight a shift in the Variation field from manual, correspondence-driven models to highly automated, data-driven and physically meaningful representations. They collectively address the need for robust morphometric analysis, cross-modal shape understanding, and patient-specific modeling. However, substantial challenges remain. Interoperability between point clouds and mesh-based representations, handling real-world heterogeneity and organ long-tail distributions, achieving interpretable and generalizable latent shape spaces, and bridging the gap to clinical utility in downstream applications are all ongoing research directions.

In summary, research on Variation in medical point cloud learning is increasingly characterized by a blend of explicit statistical models, implicit deep representations, and physiologically informed learning, paving the way toward automated, robust, and clinically meaningful anatomical shape analysis.

\begin{table}[htbp]
\centering
\caption{Overview of specialized 3D anatomical datasets}
\label{tab:dataset}
\begin{tabular}{c|c}
\hline
\textbf{Anatomical Structure} & \textbf{Dataset}                      \\ \hline
\multirow{2}{*}{Whole Body}   & Open 3D Model\cite{Vereecke2023Open}                            \\ \cline{2-2} 
                              & BodyParts3D / Anatomography\cite{mitsuhashi2009bodyparts3d}              \\ \hline
\multirow{2}{*}{Multi-organ}   & MedShapeNet\cite{li2025medshapenet}                            \\ \cline{2-2} 
                              & MedPointS\cite{zhang2025hierarchical}              \\ \hline
\multirow{2}{*}{Heart}   & PointCHD\cite{yang2020intra}                            \\ \cline{2-2} 
                              & Manifold Left/Right Ventricular Database\cite{verhulsdonk2023shape}              \\ \hline
Rib                           & RibSeg v2\cite{yang2021ribseg}                                \\ \hline
Intracranial Artery           & IntrA\cite{yang2020intra}                                    \\ \hline
Brain Cortical Regions        & AxonEM\cite{wei2021axonem}                                   \\ \hline
Pulmonary Tree                & Pulmonary Tree Labeling\cite{xie2025efficient}                  \\ \hline
\end{tabular}
\end{table}
\section{Datasets and Metrics}
\subsection{Datasets}

\begin{table*}[htbp]
\centering
\caption{Overview of shape learning-based Metrics}
\label{tab:mertics}
\resizebox{\textwidth}{!}{%
\begin{tabular}{c|c|c|c}
\hline
\textbf{Name}                     & \textbf{Formula}                                                                                                    & \textbf{Function}                                                                                                                       & \textbf{Used by Papers}                                                                                                                                                                                                                                                                                                                                                                                      \\ \hline
Chamfer Distance                  & $CD(A, B) = \sum_{a \in A} \min_{b \in B} ||a - b||^2 + \sum_{b \in B} \min_{a \in A} ||b - a||^2$                  & \begin{tabular}[c]{@{}c@{}}Measures geometric distance \\ between two point sets.\end{tabular}                                          & \begin{tabular}[c]{@{}c@{}}\cite{adams2023can,adams2024point2ssm++,adams2024weakly,beetz2023multi,beetz2024modeling,chen2021shape,chen2024cartilage,hu2021point,hu2022srt}\\ \cite{iyer2023mesh2ssm,jiang2023cpaconv,liu2025lcnet,mazzocchetti2024neural,verhulsdonk2023shape,yassin2024medshapenet,zhang2025hierarchical,zhang2025pussmpointcloudupsampling}\end{tabular} \\ \hline
Hausdorff Distance                & $HD(A, B) = \max \left\{ \sup_{a \in A} \inf_{b \in B} ||a - b||, \sup_{b \in B} \inf_{a \in A} ||b - a|| \right\}$ & \begin{tabular}[c]{@{}c@{}}Measures the largest distance from a \\ point in one set to the closest point in the other set.\end{tabular} & \cite{beetz2022point2mesh,chen2021shape,liu2025lcnet,verhulsdonk2023shape,zhang2025pussmpointcloudupsampling}                                                                                                                                                                                                                                                                               \\ \hline
Mean Absolute Error               & $MAE = \frac{1}{N} \sum_{i=1}^{N} |x_i - y_i|$                                                                      & \begin{tabular}[c]{@{}c@{}}Measures absolute \\ intensity differences.\end{tabular}                                                                                                & \cite{liu2025lcnet}                                                                                                                                                                                                                                                                                                                                                                         \\ \hline
Root Mean Squared Error           & $RMSE = \sqrt{ \frac{1}{N} \sum_{i=1}^{N} (x_i - y_i)^2 }$                                                          & \begin{tabular}[c]{@{}c@{}}Square root of mean squared error, \\ penalizes large errors more.\end{tabular}                              & \cite{liu2025lcnet,su20213d,wang2023bone}                                                                                                                                                                                                                                                                                                                                                   \\ \hline
Target Registration Error         & $TRE = \sqrt{ \frac{1}{N} \sum_{i=1}^{N} ||p_i^{pred} - p_i^{gt}||^2 }$                                             & \begin{tabular}[c]{@{}c@{}}Measures the average distance \\ between corresponding registered landmarks .\end{tabular}                   & \cite{falta2024unleashing,hansen2021deep,heinrich2024pointvoxelformer,heiselman2024image,joutard2022multi,liu2025lcnet,mekhzoum2024towards,zhang2023volumetric,zhang2024deformation}                                                                                                                                                                                                        \\ \hline
Dice Score                        & $Dice = \frac{2|A \cap B|}{|A| + |B|}$                                                                              & \begin{tabular}[c]{@{}c@{}}Measures overlap between \\ two voxels.\end{tabular}                                           & \cite{falta2024unleashing,heinrich2024pointvoxelformer,heiselman2024image,joutard2022multi,wang2023bone,xiong2022automatic,xzhang2023anatomical,xzhang2024robust,zhang2023volumetric}                                                                                                                                                                                                       \\ \hline                                        
Accuracy                          & $Accuracy = \frac{1}{|P_{GT}|} \sum_{x \in P_{GT}} \min_{y \in P_{Rec}} ||x - y||_2$                                & \begin{tabular}[c]{@{}c@{}}Measures average distance from \\ ground truth points to reconstructed points.\end{tabular}                  & \cite{bastico2024coupled,heinrich2024pointvoxelformer,mazzocchetti2024neural,xzhang2023anatomical,xzhang2024robust,zhao2023pointneuron}                                                                                                                                                                                                                                                     \\ \hline
Completeness                      & $Completeness = \frac{1}{|P_{Rec}|} \sum_{x \in P_{Rec}} \min_{y \in P_{GT}} ||x - y||_2$                           & \begin{tabular}[c]{@{}c@{}}Measures how well the reconstructed shape \\ covers the target surface ground truth.\end{tabular}                       & \cite{mazzocchetti2024neural,zhao2023pointneuron}                                                                                                                                                                                                                                                                                                                                           \\ \hline
Earth Mover's Distance            & $EMD(A, B) = \min_{\phi: A \to B} \sum_{a \in A} ||a - \phi(a)||$                                                   & \begin{tabular}[c]{@{}c@{}}Measures the minimal cost to transform\\  one distribution into another.\end{tabular}                        & \cite{chen2021shape,hu2021point,hu2022srt,mazzocchetti2024neural,zhang2025hierarchical}                                                                                                                                                                                                                                                                                                     \\ \hline
F Score                           & $F = 2 \cdot \frac{Accuracy \cdot Completeness }{Accuracy + Completeness }$                                                     & \begin{tabular}[c]{@{}c@{}}Harmonic mean of Accuracy and Completeness \\ in point-based reconstruction accuracy.\end{tabular}                & \cite{chen2024cartilage,jiang2023cpaconv,mazzocchetti2024neural,yassin2024medshapenet,zhang2025pussmpointcloudupsampling,zhao2023pointneuron}                                                                                                                                                                                                                                               \\ \hline
Point-to-Surface Distance         & $P2S = \frac{1}{|P|} \sum_{p \in P} \min_{s \in S} ||p - s||$                                                       & \begin{tabular}[c]{@{}c@{}}Measures distance from predicted points \\ to the ground truth surface.\end{tabular}                         & \cite{adams2023can,adams2024point2ssm++,adams2024weakly,verhulsdonk2023shape,zhang2025pussmpointcloudupsampling}                                                                                                                                                                                                                                                                            \\ \hline
Point-to-Point Distance           & $P2P = \frac{1}{N} \sum_{i=1}^{N} ||x_i - y_i||$                                                                    & \begin{tabular}[c]{@{}c@{}}Measures average Euclidean distance \\ between corresponding points.\end{tabular}                            & \cite{chen2021shape,hu2021point,hu2022srt}                                                                                                                                                                                                                                                                                                                                                  \\ \hline
Surface-to-Surface Distance       & $S2S = \frac{1}{|S|} \sum_{s \in S} \min_{s' \in S'} ||s - s'||$                                                    & \begin{tabular}[c]{@{}c@{}}Measures surface similarity \\ between two 3D meshes.\end{tabular}                                           & \cite{adams2024point2ssm++,iyer2023mesh2ssm,xiong2022automatic,xzhang2023anatomical,xzhang2024robust}                                                                                                                                                                                                                                                                                       \\ \hline
Normal Consistency                & $NC = \frac{1}{N} \sum_{i=1}^{N} |n_i^{pred} \cdot n_i^{gt}|$                                                       & \begin{tabular}[c]{@{}c@{}}Evaluates alignment of predicted \\ and ground-truth surface normals.\end{tabular}                           & \cite{jiang2023cpaconv,zhang2025pussmpointcloudupsampling}                                                                                                                                                                                                                                                                                                                                  \\ \hline                             
SSM - Compactness                 & $C = \sum_{i=1}^{k} \lambda_i$                                                                                      & \begin{tabular}[c]{@{}c@{}}Measures how well the shape model \\ captures variance with few components.\end{tabular}                     & \cite{adams2023can,adams2024point2ssm++,adams2024weakly}                                                                                                                                                                                                                                                                                                                                    \\ \hline
SSM - Generalization              & $\frac{1}{N} \sum_{i=1}^{N} ||x_i^{recon} - x_i^{true}||$                                                           & \begin{tabular}[c]{@{}c@{}}Measures the model \\ reconstructs unseen shapes.\end{tabular}                                                                                          & \cite{adams2023can,adams2024point2ssm++,adams2024weakly}                                                                                                                                                                                                                                                                                                                                    \\ \hline
SSM - Specificity                 & $\frac{1}{M} \sum_{j=1}^{M} \min_i ||x_j^{gen} - x_i^{true}||$                                                      & \begin{tabular}[c]{@{}c@{}}Measures whether the generated shapes\\  are similar to real examples.\end{tabular}                          & \cite{adams2023can,adams2024point2ssm++,adams2024weakly}                                                                                                                                                                                                                                                                                                                                    \\ \hline
Mapping Error                     & $ME = \frac{1}{N} \sum_{i=1}^{N} Var(\text{local neighborhoods})$                                                   & \begin{tabular}[c]{@{}c@{}}Measures correspondence consistency \\ across predicted shapes.\end{tabular}                                 & \cite{adams2024point2ssm++,adams2024weakly}                                                                                                                                                                                                                                                                                                                                                 \\ \hline
Structure Average          & $ESA = \frac{1}{N} \sum_{i=1}^{N} Error_i$                                                                          & \begin{tabular}[c]{@{}c@{}}Mean error over all points or \\ voxels in structure.\end{tabular}                                    & \cite{zhao2023pointneuron}                                                                                                                                                                                                                                                                                                                  \\ \hline
Jacobian Determinant              & $J = det\left(\frac{\partial f}{\partial x}\right)$                                                                 & \begin{tabular}[c]{@{}c@{}}Evaluates local volume \\ change in deformation fields.\end{tabular}                                         & \cite{heiselman2024image,zhang2023volumetric}                                                                                                                                                                                                                                                                                                                                               \\ \hline
\end{tabular}%
}
\end{table*}

In 3D point‑cloud–based medical shape analysis, publicly available datasets play a pivotal role\cite{li2025medshapenet}. A diverse collection of anatomical shapes enables the development and evaluation of data‑driven, shape‑based methods tailored to both visual and clinical challenges. Public datasets furnish researchers with a common benchmark, allowing methods to be evaluated under consistent conditions and thereby driving algorithmic optimization and advancement\cite{ogier2022flamby}. Although existing 3D object datasets such as ShapeNet\cite{chang2015shapenet} contain CAD models of real‑world objects (e.g. airplanes, cars, chairs, tables), they lack sufficient anatomical diversity. Prior to the release of MedShapeNet, most point‑cloud–based medical shapes were extracted from 3D masks derived from volumetric medical images.

We can distinguish two typical processing approaches, exemplified by MR‑Net\cite{chen2021shape} and PCD‑Net\cite{beetz2024modeling}, both of which utilize cardiovascular MRI from the UK Biobank\cite{petersen2016uk}. MR‑Net converts the voxelized exterior contour into point clouds, thereby capturing features of the immediate surface layer. However, voxel‑based representations—unlike CAD‑derived 3D objects—do not emphasize shape characteristics such as jagged edges, volume, or elongation. When converting voxels into point clouds, the resulting representation is often overly sparse and dispersed, failing to accurately reflect fine shape details. PCD‑Net, in contrast, reconstructs a mesh from the voxel surface before sampling point clouds from that mesh; this method better preserves surface‑level shape features. Nevertheless, unless mesh artifacts are explicitly addressed, errors such as incorrect connections or non‑manifold structures may arise. 

The IntrA dataset\cite{yang2020intra} for 3D cardiovascular modeling exemplifies a rigorous processing pipeline: segmentation of cardiovascular structures, surface reconstruction, cleanup of non‑manifold artifacts, and generation of a valid 3D model. This standard pipeline underpins why we emphasize at the outset that segmentation is foundational to Registration, Reconstruction, and Variation. Likewise, MedShapeNet provides accurately processed 3D anatomical shapes derived from real patient imaging data, including both healthy and pathological cases.

To this end, we catalog existing specialized 3D anatomical datasets: precisely curated anatomical shapes significantly aid in the development and evaluation of methods for visual and medical problems, as summarized in Table \ref{tab:dataset}. Considering the long‑tail distribution of anatomical variants in medical imaging, the development of dedicated datasets for medical shape analysis and 3D point‑cloud–based shape learning remains essential.

\subsection{Metrics}

Evaluation metrics play a pivotal role in 3D point-cloud–based medical shape analysis by providing standardized performance assessment, guiding algorithmic refinement, and validating clinical applicability. Scientifically rigorous and well-designed metrics offer objective evaluation of registration accuracy, reconstruction fidelity—including detail preservation and noise suppression—ensuring algorithmic utility across diverse use cases\cite{erickson2021magician}. Since different metrics target distinct aspects of point-cloud quality\cite{rainio2024evaluation}, relying on a single metric is often inadequate to fully capture the multifaceted goals of shape learning. Consequently, comprehensive and reliable assessment typically involves a combination of complementary metrics. In this review, we summarize commonly used evaluation metrics in 3D medical point-cloud shape learning—results are presented in Table \ref{tab:mertics}.

Unsurprisingly, the most frequently used metrics are Target Registration Error (TRE) for registration tasks and Chamfer Distance (CD) for reconstruction tasks. In the context of registration, the Dice Score is also extensively employed to quantify volumetric overlap and assess overall structural alignment quality. For reconstruction tasks involving point-cloud completion, additional metrics such as Hausdorff Distance (HD) and F-score are often incorporated to capture fine-grained differences. When evaluating surface reconstruction, metrics like Normal Consistency are included to assess the accuracy of surface orientation and smoothness.

Regarding variation modeling, beyond Chamfer Distance, other geometric discrepancy metrics such as point-to-point (P2P), point-to-surface (P2S), and surface-to-surface (S2S) distances are frequently used. These point- or surface-level distances more precisely reflect anatomical differences and inter-subject variability. Moreover, analyses based on Statistical Shape Models (SSMs)\cite{iyer2023mesh2ssm,adams2024point2ssm++,adams2024weakly,adams2023can} emphasize metrics such as compactness, generalization, and specificity to evaluate the modeling capacity and practical relevance of SSMs, particularly in terms of their ability to represent, generate, and generalize anatomical structures. In biomechanics-informed frameworks, Jacobian determinant–based measures are also employed to quantify local volumetric deformation.

\section{Summary and Future Directions}

Medical point cloud shape learning has seen rapid advances in recent years, driven by progress in deep learning, hybrid representations, and the availability of large-scale datasets. This survey systematically reviewed state-of-the-art approaches in registration, reconstruction, and variation modeling, as well as their clinical relevance, datasets, and evaluation criteria.

Despite notable progress, several challenges remain. Annotated medical point cloud data is still limited, hindering the training and validation of robust models—especially for rare diseases, pediatric cases, and complex anatomical variations. Cross-modal generalization, model interpretability, and integration into real-world clinical workflows also require further research. In addition, current benchmarks may not fully capture the nuanced requirements of different clinical applications.

Looking ahead, future work should focus on: 1. developing advanced generative and self-supervised learning strategies to leverage unlabeled data and reduce annotation costs; 2. improving model generalizability and fairness across diverse populations and imaging modalities; 3. designing interpretable, trustworthy, and clinically usable shape learning models; 4. establishing new datasets and clinically meaningful benchmarks tailored to medical point cloud tasks. Integrating point cloud-based shape learning with multimodal, temporal, or physiological information will further expand the boundaries of medical image analysis, supporting next-generation intelligent healthcare systems.

\bibliographystyle{IEEEtran}
\bibliography{reference}


\end{document}